# Generalizability of PRS$_{313}$ for breast cancer risk amongst non-Europeans in a Los Angeles biobank

**Authors**
Helen Shang*[1,2], Yi Ding[3], Vidhya Venkateswaran[4], Kristin Boulier[5], Nikhita Kathuria-Prakash[6], Parisa Boodaghi Malidarreh[7,8], Jacob M. Luber,[7,8,9] Bogdan Pasaniuc,[10,11,12,13,14]

**Affiliations**
[1] Division of Internal Medicine, Ronald Reagan UCLA Medical Center, Los Angeles, CA, USA.
[2] Department of Computer Science and Engineering, University of Texas at Arlington, Arlington, TX, USA.
[3] Bioinformatics Interdepartmental Program, University of California, Los Angeles (UCLA), Los Angeles, CA, USA.
[4] Department of Oral Biology, UCLA School of Dentistry, Los Angeles, CA, USA.
[5] Division of Cardiology, Department of Medicine, Ronald Reagan UCLA Medical Center, Los Angeles, CA, USA.
[6] Division of Hematology-Oncology, Department of Medicine, Ronald Reagan UCLA Medical Center, Los Angeles, California, USA.
[7] Department of Computer Science and Engineering, University of Texas at Arlington, Arlington, Texas, USA.
[8] Multi-Interprofessional Center for Health Informatics, University of Texas at Arlington, Arlington, Texas, USA.
[9] Department of Bioengineering, University of Texas at Arlington, Arlington, Texas, USA.
[10] Bioinformatics Interdepartmental Program, University of California, Los Angeles (UCLA), Los Angeles, CA, USA.
[11] Department of Pathology and Laboratory Medicine, David Geffen School of Medicine, University of California, Los Angeles, Los Angeles, CA, USA.
[12] Department of Human Genetics, David Geffen School of Medicine, University of California, Los Angeles, Los Angeles, CA, USA.

[13] Department of Computational Medicine, David Geffen School of Medicine, University of California, Los Angeles, Los Angeles, CA, USA.
[14] Institute of Precision Health, University of California, Los Angeles, Los Angeles, CA, USA.

*Correspondence: hshang@mednet.ucla.edu

**Abstract**

Polygenic risk scores (PRS) summarize the combined effect of common risk variants and are associated with breast cancer risk in patients without identifiable monogenic risk factors. One of the most well-validated PRSs in breast cancer to date is PRS313, which was developed from a Northern European biobank but has shown attenuated performance in non-European ancestries. We further investigate the generalizability of the PRS313 for American women of European (EA), African (AFR), Asian (EAA), and Latinx (HL) ancestry within one institution with a singular EHR system, genotyping platform, and quality control process. We found that the PRS313 achieved overlapping Areas under the ROC Curve (AUCs) in females of Lantix (AUC, 0.68; 95 CI, 0.65-0.71) and European ancestry (AUC, 0.70; 95 CI, 0.69-0.71) but lower AUCs for the AFR and EAA populations (AFR: AUC, 0.61; 95 CI, 0.56-0.65; EAA: AUC, 0.64; 95 CI, 0.60-0.680). While PRS313 is associated with Hormone Positive (HR+) disease in European Americans (OR, 1.42; 95 CI, 1.16-1.64), for Latinx females, it may be instead associated with Human Epidermal Growth Factor Receptor 2 (HER2+) disease (OR, 2.52; 95 CI, 1.35-4.70) although due to small numbers, additional studies are needed. In summary, we found that PRS313 was significantly associated with breast cancer but with attenuated accuracy in women of African and Asian descent within a singular health system in Los Angeles. Our work further highlights the need for additional validation in diverse cohorts prior to clinical implementation of polygenic risk scores.

**Introduction**

The US Preventive Services Task Force (USPSTF) recommends that breast cancer screening start at 50 years old, based on studies showing that 90% of breast cancer cases are diagnosed after this age[1]. Unfortunately, this also means that 10% of cases will be missed per conventional guidelines, equating to approximately 10,000 missed cases in the US annually[2]. As such, researchers have been working to develop new methods of identifying patients at risk of developing early-onset breast cancer. One well-known approach is the Gail model, which uses clinical, family history, and demographic information to calculate individual breast cancer risk but suffers from poor accuracy; in a meta-analysis across 26 studies and 29 datasets, a modified version of the Gail model had AUCs for American, Asian and European females of 0.61 (95 CI 0.59–0.63), 0.55 (95 CI 0.52–0.58) and 0.58 (95 CI 0.55–0.62), respectively[3,4]. Another avenue has been identifying carriers of genes such as *BRCA1*, *BRCA2*, *PTEN*, *TP53*, *CDH1*, and *STK11*, which are associated with a two-to-threefold elevated risk in developing breast cancer over a patient's lifetime[5]. However, the vast majority (>75%) of breast cancer patients do not have any identifiable monogenic risk factors and thus will not benefit from this approach. Over the past decade, polygenic risk scores (PRSs) were introduced as a potential solution by summarizing the combined effect of multiple common risk variants that have been individually associated with small, yet elevated breast cancer risk[6]. Several PRSs have been developed to predict and stratify breast cancer risk, including a recent paper showing that at the top 5th percentile of one PRS had genetic risk of similar magnitude to some monogenic etiologies[7]. Prior work has estimated that the theoretically best PRS, if the effect sizes of all common SNPs were known with certainty, would explain ~41% of the familial risk of breast cancer[8].

One of the most validated PRSs in breast cancer to date is the $PRS_{313}$ by Mavaddat et al., which was developed from a Northern European biobank (n = 33,673 cases and n = 33,381 controls). When incorporating family history and age of diagnosis, the $PRS_{313}$ achieved an OR of 1.61 (95 CI, 1.57 to 1.65) and AUC of 0.63 (95 CI = 0.63-0.65)[9]. Subsequent work has demonstrated an attenuated effect of $PRS_{313}$ in

African females; for example, Cong et al. found that for 33,594 women of European ancestry and 2,801 women of African ancestry across 9 institutions, the $PRS_{313}$ alone achieved a higher AUC for European females (0.60, 95 CI, 0.59-0.61) than for African females (0.55, 95 CI, 0.51-0.58)[10,11].

In this paper, we aim to further investigate the generalizability of the $PRS_{313}$ for American women of African, Asian, and Latinx ancestry within one institution, leveraging a singular EHR system, genotyping platform, and quality control process.

**Material and Methods**

**Study Participants**

The participants included in this cohort study were females at birth drawn from the UCLA ATLAS Biobank (N=18,627), which is linked to electronic medical record (EMR) data and has been described previously[12]. SNPs were genotyped on a genome-wide array and imputed to the TOPmed reference panel. We identified breast cancer cases and controls using ICD-9 and ICD-10 codes corresponding to Phecode X 105.1, specifying "Malignant neoplasm of the breast, female," which maps ICD codes to clinically meaningful phenotypes [12].

We previously identified five Genetically Inferred Ancestries (GIA) based on Principal Component Analysis (PCA) and k-means clustering including African Americans (AA), Hispanic Latino Americans (HL), East Asian Americans (EAA), European Americans (EA), and South Asian American (SAA)[13]. The SAA population was not included for further analysis due to case counts being significantly underpowered. As we did not have access to individual-level pathology results, we identified hormone receptor positive (HR+) or human epidermal growth factor receptor 2 positive (HER2+) breast cancer based on prescription data correlating to either subtype.

**PRS Models**

The PRS was calculated based on an additive model using effect size estimates from the polygenic score as initially developed by Mavaddat et al., which are available under the entry PGS000004 within the The Polygenic Score (PGS) Catalog[14]:

Equation 1:
$$PRS = \beta_1 x_1 + \beta_2 x_2 + \ldots \beta_k x_k \ldots + \beta_n x_n$$

Where $\beta_k$ is the per-allele log odds ratio (OR) for breast cancer associated with the minor allele for SNP $k$, and $x_k$ the number of minor alleles for the same SNP (0, 1, or 2), and n = 313 is the total number of SNPs. 30 SNPs were excluded due to ambiguity and 45 other SNPs were unmatched. As with the original study, the raw PRS of the European population was normalized to a standard deviation of 1 and mean of 0. We then applied normalization using the average and standard deviation of European samples to the remaining GIAs as done in prior studies testing the generalizability of $PRS_{313}$ [10,11].

## Genotyping

Details of genotyping, imputation, and quality control procedures of our cohort have been previously described [13]. For this study, variants that match the following 3 criteria were retained for PRS calculation: (1) a mean R2 imputation quality greater than 0.3 across genotype array-batches; (2) P value greater than $1 \times 10^{-6}$ in ancestry-specific Hardy Weinberg Equilibrium tests; and (3) minor allele frequency (MAF) greater than 0.005. We then performed LD pruning with plink2 (--indep-pairwise 1000 50 0.05) and excluded the long-range LD regions. The top nine PCs were computed with the flashpca2 software[15].

## Statistical Methods

Logistic regression models were used to estimate the odds ratios (ORs) for the PRS on breast cancer with age and the first nine principal components as covariates using the equation below:

Equation 2:

$$\log(\text{Breast cancer}) = \beta_0 + \beta_1(\text{PRS}) + \beta_2(\text{Age}) + \beta_3(\text{PC1}) + \ldots + \beta_{11}(\text{PC9})$$

Rather than using age of diagnoses as per Mavaddat et al., we used the age at which an ICD9 or ICD10 corresponding to breast cancer appeared in the patient's medical record. This is due to the fact that the age of diagnosis was not stored as structured data from our EHR. We were also unable to include family history as a covariate for the same reason. To be consistent with Mavaddat et al., we also included the first nine principal components as covariates to account for potential differences in population structure across ancestries.

Logistic regression with the same covariates were used to estimate the ORs for breast cancer by deciles of the PRS, with the middle (50th percentile) as the reference. Percentiles and their ORs were calculated per GIA separately. To examine the discrimination of each PRS per GIA, we estimated the area under the receiver operator characteristic curves (AUC) using the standardized PRS score, age at biobanking, and first nine principal components as predictors.

**Survival Analysis**

Kaplan Meier analysis was performed by using the PRS as a sole predictor for Overall survival (OS) in days as a continuous variable. OS was calculated by subtracting death or present time from the first date at which an ICD9 or ICD10 code corresponding to breast cancer appeared in a patient's medical record. Patients without OS values were discarded from the analysis (N=12).

**Results**

Our study included 18,627 women, including 1,156 with African ancestry (AA), 11,873 women with European ancestry (EA), 2,010 with East Asian ancestry (EAA), and 3,303 with Hispanic ancestry (HL) as presented in Table 1. The majority of cases were identified as HR+ and the prevalence across ancestries was comparable to the latest SEER registry showing 70% breast cancer subtypes are HR+[16]. Similarly, we found prevalences of HER2+ disease that were also comparable to SEER registries showing a prevalence of 10.8%. For 1285 out of 2080 cases, we did not have any prescription data as these patients may have undergone cancer treatment outside of the UCLA network.

**Association of $PRS_{313}$ With Breast Cancer Risk in Various Ancestries**

For each GIA group, the PRS followed a normalized distribution with the EAA cohort having a higher mean, as illustrated in Figure 1 (AA: N= 1156; mean: -0.02; std: 0.46; EA: N= 11,873; mean: -0.06; std: 0.53; EAA: N= 2010; mean: 0.15; std: 0.47; HL: N= 3,303; mean: -0.06; std: 0.53). We found statistically significant associations of $PRS_{313}$ with overall breast cancer risk across all ancestries. All GIAs had overlapping ORs (AA: OR, 1.31; 95 CI, 1.05-1.64; EA: OR, 1.52; 95 CI, 1.23-1.72; EAA: OR, 1.46; 95 CI, 1.23-1.61; HL: OR, 1.51, 95 CI, 1.31-1.75). These also overlap with Mavaddat et al.'s reported OR in Europeans (OR, 1.61; 95 CI, 1.57 to 1.65).

For all GIAs, the ORs of the $PRS_{313}$ was largest at the extremes PRS distribution relative to the 50th percentile and the 95% confidence intervals overlap among the four GIAs (AA: OR, 1.85; 95 CI, 1.08-3.2; EA: OR, 2.1; 95 CI, 1.9-2.6; EAA: OR, 1.7; 95 CI, 1.1-2.6; HL: OR, 1.7; 95 CI, 1.1-2.5: ). For the AA population, we found an attenuated effect of the PRS and at the extremes of its distribution (>95 percentile) relative to other GIAs, although this analysis was limited by small sample sizes (not shown).

**Discriminative accuracy of the $PRS_{313}$**

The discriminative accuracy of the PRS for any type of breast cancer, as measured by the AUC was highest in the EA population at 0.70 (95 CI = 0.69 to 0.71), which is slightly higher than 0.63 (95 CI = 0.63 to 0.65) as reported by Mavaddat et al. Relative to the EA population, the AUC for the HL population was similar whereas those for the AFR and EAA populations were lower (AFR: AUC, 0.61; 95 CI, 0.56-0.65; EAA: AUC, 0.64; 95 CI, 0.60-0.680; HL: AUC, 0.68; 95 CI, 0.65 -0.71).

**Association of $PRS_{313}$ With Breast Cancer Subtypes in Various Ancestries**

Recent work in European females has shown that HR+ breast cancer is associated with $PRS_{313}$ and a lower risk genetic signature[17]. As with prior work, we found that for European patients, the $PRS_{313}$ was associated with HR+ disease; however, it was not associated with HR+ disease in Hispanic, African, nor Asian American females with breast cancer (Table 2). In contrast, amongst Hispanic Americans alone, HER2+ disease was instead associated with $PRS_{313}$ with a higher OR than for all breast cancer risk (OR, 2.47; 95 CI, 1.39 - 4.41) although this analysis was limited as there were only 18 patients within our cohort that were both Hispanic and HER2+. We also noted that the average age of diagnosis amongst HER2+ Hispanic patients was notably lower than other ethnicities (AA mean: 60 (SD: 15); EA mean: 56 (SD: 14); EAA mean: 51 (SD: 11); HL mean: 44 (SD: 7.7)) and we therefore investigated whether age at diagnosis might confounding these results; in their original work, Mavaddat et al. showed a decline in relative risk with age and $PRS_{313}$. However, the association of $PRS_{313}$ with HER2+ disease remained even when accounting for age at diagnosis (OR, 2.52; 95 CI, 1.35-4.70).

**Survival analysis**

To confirm that our estimates of OS were reliable, we first confirmed that OS is appropriately associated with whether or not a patient had received chemotherapy, as a hallmark of more aggressive and often metastatic disease (Supplementary Figure 1). We found that in European patients, the $PRS_{313}$ alone as a predictor fails to stratify patients by survival time by Kaplan Meier analysis when comparing breast cancer patients above (N=280) or below (N=651) the 70th percentile with a log-rank p-value of 0.38 (Supplementary Figure 2a). We chose the 70th percentile arbitrarily as we wanted to select a threshold at which the OR of $PRS_{313}$ would be higher than the 50th percentile, as illustrated in Figure 1. Of note, we found similar results when evaluating different thresholds, such as comparing the top vs bottom 50th percentiles and top 90th vs bottom 10th percentiles. For European patients, after accounting for whether a patient had received chemo, age of diagnosis, HER2+, and PR+ disease, $PRS_{313}$ was also no longer predictive of OS, using a multivariate Cox proportional hazard model (Supplementary Figure 2b). This is consistent with recent work involving a cohort of European (N=98,397) and Asian (N=12,920) females with breast cancer which found that in European patients, $PRS_{313}$ was no longer associated with OS after adjusting for breast cancer subtype and tumor grade[17].

**Discussion**

In this paper, we investigated the generalizability of the $PRS_{313}$ for American women of African, Asian, and Latinx ancestry within one institution. For females of European ancestry, we arrived at overlapping estimates of OR (1.5; 95 CI, 1.2-1.7) when compared to Mavaddat et al. (1.61; 95 CI, 1.57-1.65). Consistent with prior studies, we found that the $PRS_{313}$ was still associated with breast cancer across all ancestries but with an attenuated effect in females of African and Asian ancestry; the $PRS_{313}$ achieved equivalent AUCs in females of Latinx (AUC, 0.68; 95 CI, 0.65-0.71) and European ancestry (AUC, 0.70; 95 CI, 0.69-0.71) with lower AUCs for the AFR and EAA populations (AFR: AUC, 0.61; 95 CI, 0.56-0.65; EAA: AUC, 0.64; 95 CI, 0.60-0.680). This may be due to Latinx individuals having a greater proportion of European ancestry than Asians and Africans; previously, we found that Latinx individuals in the UCLA ATLAS cohort have a complex genetic admixture with its PCA substructure centered on the European samples with arms extending into the African and Asian groups[13].

Consistent with prior work, we found that for European American patients, $PRS_{313}$ is associated with HR+ disease[9,17]. For African, Hispanic, and Asian Americans, this association is lost although rates of HR+ disease were lower in these cohorts relative to European Americans. Surprisingly, we found that for Hispanic Americans alone, the $PRS_{313}$ was associated with HER2+ disease, which has not been previously reported. These findings cannot be entirely explained by a higher prevalence of HER2+ disease amongst Hispanic patients as we found that within our cohort, Hispanics had lower or equivalent rates of HER2+ disease relative to other ethnicities. Furthermore, this association remained true even after accounting for age at diagnosis. We noticed Hispanics were diagnosed with breast cancer at younger years relative to other ethnicities, which has been shown previously, and in their original paper, Mavaddat et al. showed a decline in relative risk with age and $PRS_{313}$[18]. However, given our cohort's limited size, further investigation regarding this association is warranted.

Similarly, as with recent work on $PRS_{313}$, we found that it fails to stratify European patients by OS[15]. One possible explanation is that the impact of $PRS_{313}$ on OS may be confounded by other genetic risk factors, many of which have yet to be identified;

several recent papers have found that $PRS_{313}$ stratifies breast cancer risk in *CHEK2, PALB2*, and *ATM* carriers but not *BRCA1*/2 carriers[19,20]. In other words, the value of $PRS_{313}$ may be in stratifying carriers of low-penetrance risk variants but may fail to stratify those with highly penetrant variants who will go on to develop breast cancer regardless. While we were not able to confirm this in our study given the few number of risk carriers in our cohort, we hope to validate this in future studies.

There are many limitations of this study to consider. Our cohort contained fewer non-European than European samples and thus, analyses at the upper extremes and within subtypes were limited. We also could not estimate the absolute risk of developing breast cancer due to the lack of longitudinal outcomes data. Furthermore, many covariates such as age of diagnoses, cancer subtype, overall survival were not available in the medical record as structured data and were thus calculated by proxy methods. Nevertheless, we were able to confirm that these estimates resulted in expected observations suggesting their reliability, such as the expected prevalence rates amongst breast cancer subtypes and relationship between chemotherapy as a predictor of OS.

In summary, we found that $PRS_{313}$ was significantly associated with breast cancer in American females of diverse ancestries but with attenuated accuracy in women of African and Asian descent within a singular yet diverse biobank in Los Angeles. While the $PRS_{313}$ is associated with HR+ disease in European Americans, this association is lost in African, Hispanic, and Asian Americans. For Hispanic Americans, $PRS_{313}$ may be instead associated with HER2+ disease although due to small numbers, additional studies will be critical in validating these findings. Our work further highlights the need for additional validation in diverse cohorts prior to clinical implementation of polygenic risk scores and the need for new methods that can address differences in genomic admixture.

**Supplemental List**

Supplemental Figure 1: Kaplan Meier Curve for Overall Survival in European Patients with or without Chemotherapy Treatment

Supplemental Figure 2: Kaplan Meier Curve for Overall Survival in European Patients above or below the 70th Percentile of $PRS_{313}$

Supplemental Figure 3: Multivariable Cox Proportional Hazards Model on OS in European Patients

Supplemental Table 1a: Medications used for identifying HR+ breast cancer cases

Supplemental Table 1b: Medications used for identifying HER2+ breast cancer cases

Supplemental Table 2: Association between PRS and Breast Cancer Risk in GIAs

**Supplemental Figure 1: Kaplan Meier Curve for Overall Survival in European Patients with or without Chemotherapy Treatment**

We approximated Overall Survival (OS) in days by subtracting the present day or date of death from date of diagnosis in patients who were still living or dead, respectively. To validate this approach, we confirmed that the approximated OS is appropriately shorter in all patients who had received chemotherapies for breast cancer relative to those who had not (p value < .005). This is expected as patients who receive chemotherapy often have more aggressive or metastatic disease, resulting in shorter survival. There were 503 patients who had received chemotherapy and 767 without chemotherapy in our cohort.

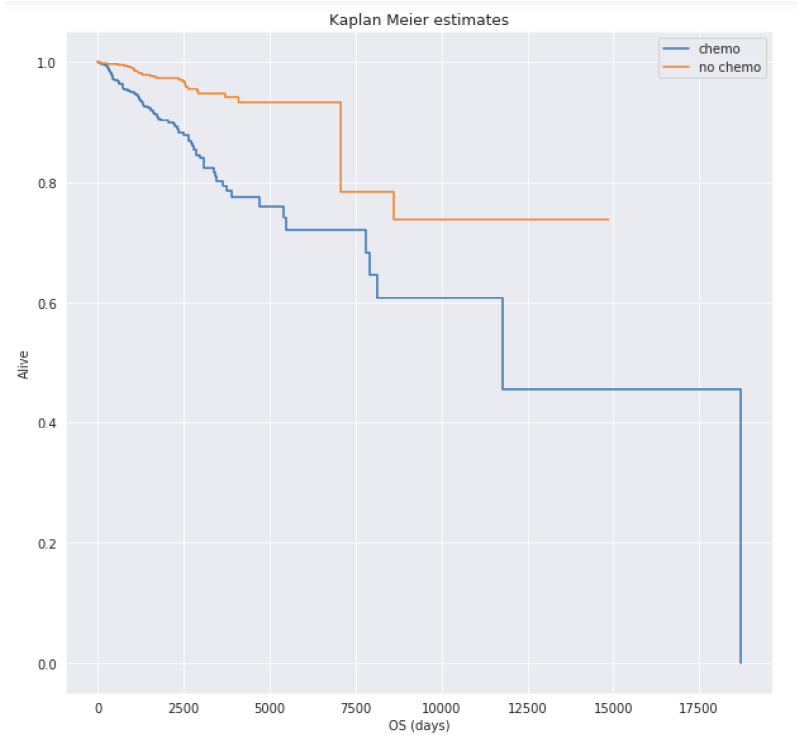

**Supplemental Figure 2: Kaplan Meier Curve for Overall Survival in European Patients above or below the 70th Percentile of $PRS_{313}$**

To evaluate the impact of $PRS_{313}$ on OS, we compared survival times by Kaplan Meier analysis for European patients above (N=280) and below (N=651) the 70th percentile of the PRS. We found no difference in survival time between the two groups by log-rank test (p-value= 0.38). There was also no difference in survival time when comparing above and below the 50th percentile as well as the top 90th and lowest 10th percentiles.

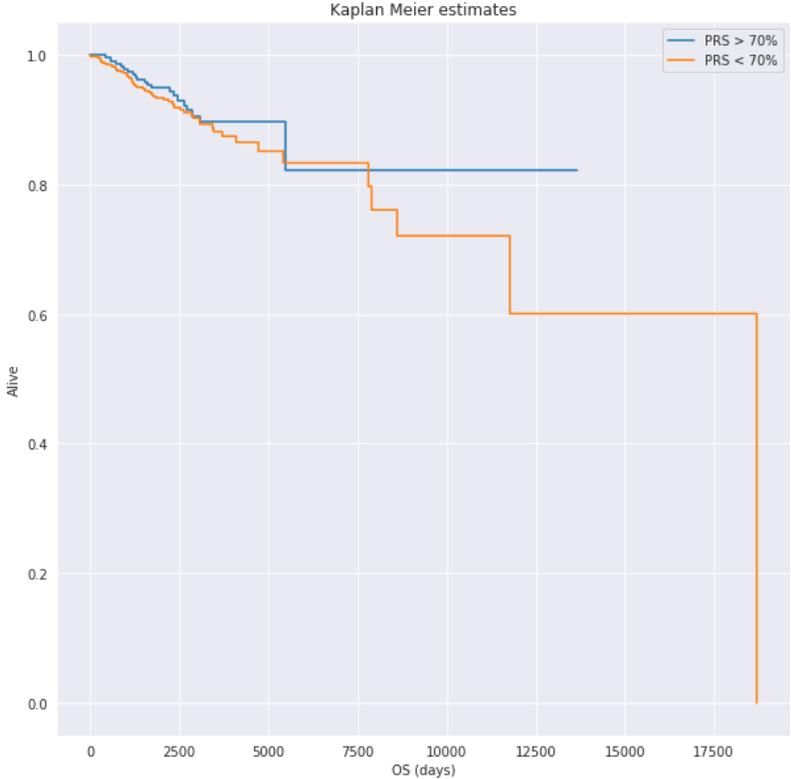

**Supplemental Figure 3: Multivariable Cox Proportional Hazards Model on OS in European Patients**

For European patients, we initially found by Cox Proportional Hazards that the normalized $PRS_{313}$ was inversely predictive of OS, suggesting that a lower $PRS_{313}$ score translates to longer survival time (HR, 0.80; 95 CI, 0.64-0.99, p-value = 0.04). However, when adjusted for other variables such as whether or not a patient had received chemotherapy, cancer subtype (HR+ and/or HER2+), and age of diagnosis, the normalized $PRS_{313}$ score is no longer predictive of OS (p–value = 0.06).

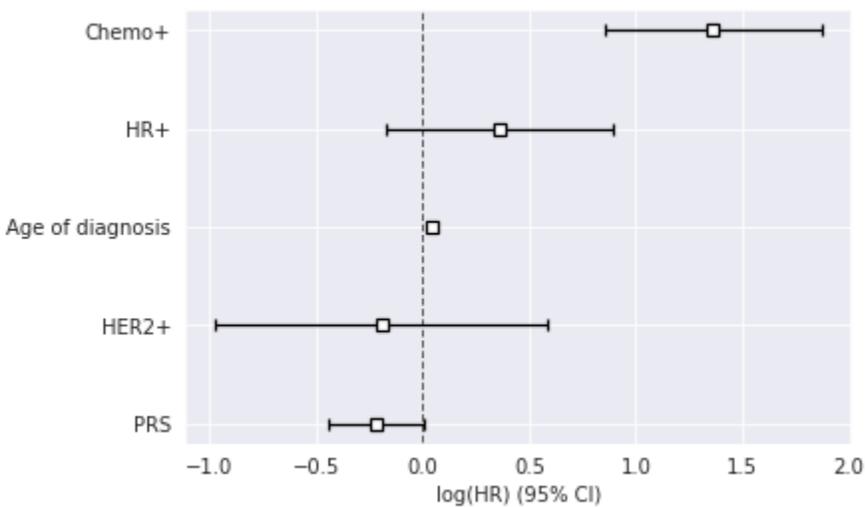

**Supplemental Table 1a: Medications used for identifying HR+ breast cancer**
**Supplemental Table 1b: Medications used for identifying HER2+ breast cancer**

We first queried the top 50 most commonly prescribed cancer-related medications within the electronic medical record for our cohort. These were manually reviewed and grouped based on class and subtype relevance. Supplemental Table 1a shows our grouped list of HR+ relevant medications and Supplemental Table 2a shows our grouped list of HER2+ relevant medications. Patients were subtyped as HR+ and/or HER2+ if they had been ordered at least one relevant medication from each list.

| Name | Medication Class |
| --- | --- |
| Exemestane | Hormone therapy |
| Anastrazole | Hormone therapy |
| Letrozole | Hormone therapy |
| Tamoxifen | Hormone therapy |
| Raloxifine | Hormone therapy |
| Fulvestrant | Hormone therapy |
| Palbociclib | CDK 4/6 inhibitor |
| Ribociclib | CDK 4/6 inhibitor |
| Abemaciclib | CDK 4/6 inhibitor |
| Everolimus | mTOR inhibitor |
| Alpelisib | PI3K inhibitor |

| Name | Medication Class |
| --- | --- |
| Trastuzumab | HER2 monoclonal antibody |
| Pertuzumab | HER2 monoclonal antibody |
| Neratinib | Tyrosine Kinase Inhibitors |
| Lapatinib | Tyrosine Kinase Inhibitors |
| Ruxolitinib | Tyrosine Kinase Inhibitors |
| Osimertinib | Tyrosine Kinase Inhibitors |
| Tucatinib | Tyrosine Kinase Inhibitors |

| Niraparib | Tyrosine Kinase Inhibitors |
| Dasatinib | Tyrosine Kinase Inhibitors |

**Supplemental Table 2: Association between PRS and Breast Cancer Risk in Genetically Inferred Ancestries (GIA)**

Univariate logistic regression was used to evaluate the association between $PRS_{313}$ and observed rates of all breast cancers in American females of African (AA), European (EA), East Asian American (EAA), and Hispanic (HL) genetically inferred ancestries (GIA). As with prior studies testing the generalizability of $PRS_{313}$, we normalized the raw PRS score of non-European GIAs based on the average and standard deviation of European samples. In comparison, Mavaddat et al. reported an OR of 1.61 (95 CI 1.57-1.65), which overlaps with all GIAs.

| GIA | OR | Lower CI | Upper CI |
| --- | --- | --- | --- |
| AA | 1.31 | 1.05 | 1.64 |
| EA | 1.52 | 1.23 | 1.72 |
| EAA | 1.46 | 1.43 | 1.61 |
| HL | 1.51 | 1.31 | 1.75 |

**Acknowledgements**

We gratefully acknowledge the Institute for Precision Health, participating patients from the UCLA ATLAS Precision Health Biobank, UCLA David Geffen School of Medicine, UCLA Clinical and Translational Science Institute, and UCLA Health. VV is funded by NIH/NIDCR 5K12DE027830-04. The ATLAS Community Health Initiative is supported by UCLA Health, the David Geffen School of Medicine, and a grant from the UCLA Clinical and Translational Science Institute (UL1TR001881). BP was partially supported by NIH awards R01 HG009120, R01 MH115676, R01 CA251555, R01 AI153827, R01 HG006399, R01 CA244670, and U01 HG011715. KB is supported by UCLA T32 funding.

**Declaration of Interests**

None

**Tables and Figures**

**Table 1: Participant Characteristics**

Participants were females drawn from the UCLA ATLAS Biobank (N=18,627), which is linked to UCLA medical records from 2013 to present day. Cases and controls were identified based on ICD-9 and ICD-10 coding corresponding to breast cancer. Age at diagnosis was based on the date at which the ICD code appeared in a patient's chart, which was then used to calculate Overall Survival (OS), with the day of death or present day as an end point. Breast cancer subtypes were identified based on prescriptions ordered.

|  | AA | EA | EAA | HL |
|---|---|---|---|---|
| Controls | 1032 | 10407 | 1803 | 3083 |
| Cases (% of total) | 124 (10) | 1466 (12) | 207 (10) | 220 (6.7) |
| Age at diagnosis, mean (SD) | 55 (17) | 55 (17) | 52 (17) | 48(17) |
| Cases with prescription data (% of total) | 86 (52) | 1241 (58) | 188 (64) | 166 (58) |
| HR+ (% of cases with prescription data) | 60 (71) | 979 (72) | 129 (65) | 115 (65) |
| HER2+ (% of cases with prescription data) | 11 (12) | 130 (10) | 26 (14) | 18 (11) |
| OS in days, mean (SD) | 2862 (1922) | 2741 (1973) | 2569 (1427) | 2911 (2149) |

# Figure 1: Distribution of PRS$_{313}$ in cases and controls

Kernel distribution estimation plots of PRS$_{313}$ scores in cases and controls by genetically inferred ancestry (GIA). The raw PRS$_{313}$ scores of the European population (EA) was normalized to a standard deviation of 1 and mean of 0. The remaining GIAs were normalized to the average and standard deviation of EA samples.

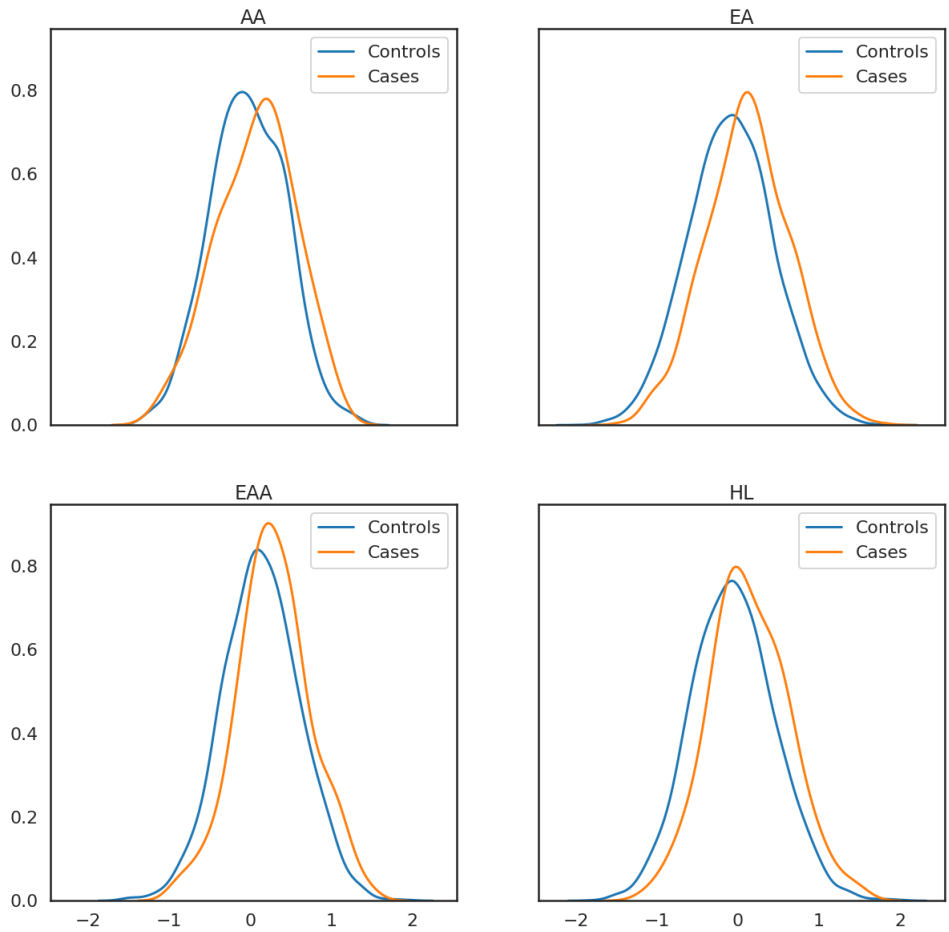

**Figure 2: Association of $PRS_{313}$ Deciles With Breast Cancer Relative to the 50th Percentile**

Association between $PRS_{313}$ and breast cancer diagnoses in American women of African (AA), European (EA), East Asian American (EAA), and Hispanic (HL), based on genetically inferred ancestries. Odds ratios and 95% confidence intervals are shown. Odds ratios are for different deciles of the PRS relative to the $50^{th}$ percentile of the PRS.

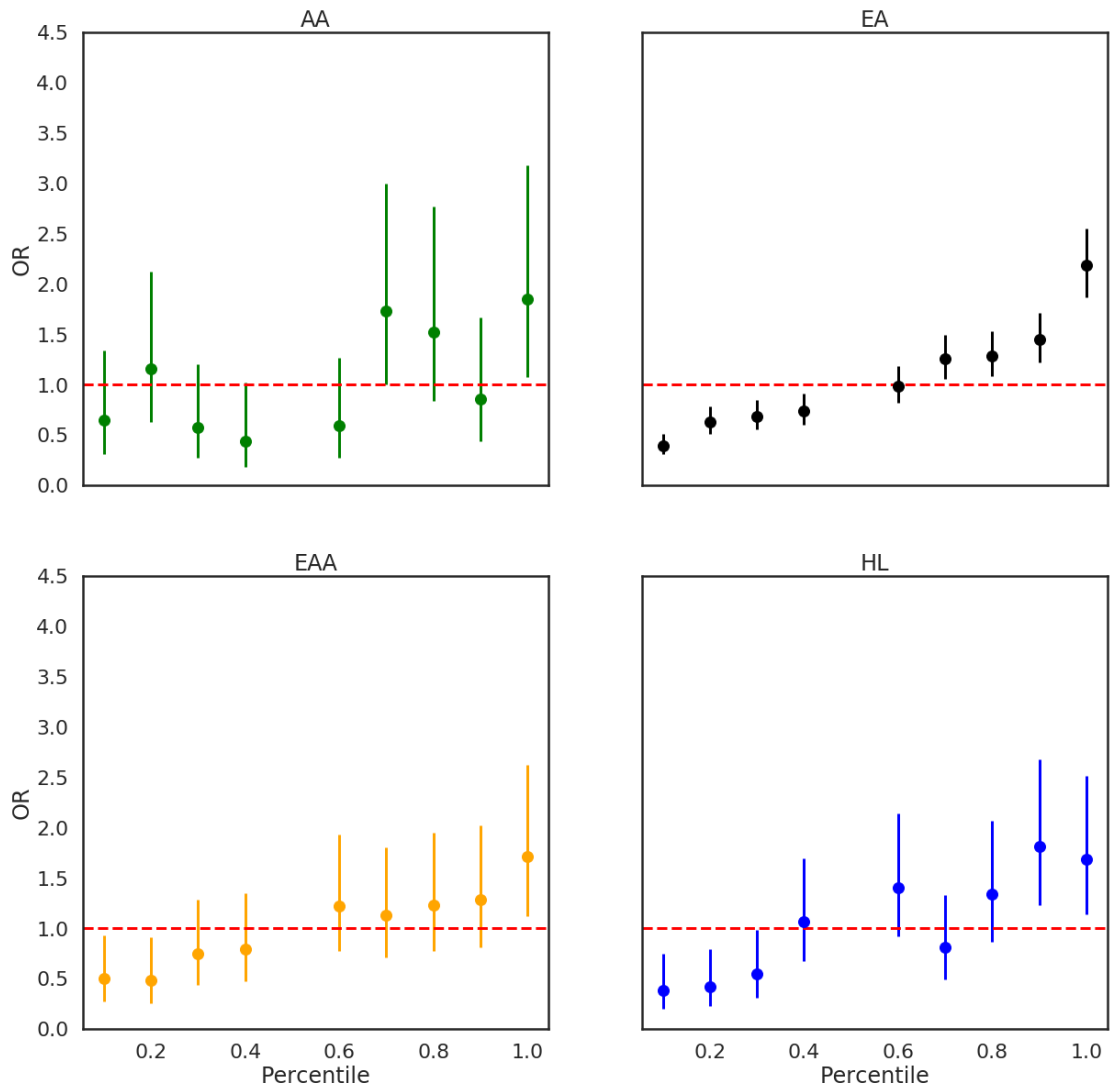

**Table 2 Association of PRS$_{313}$ With Breast Cancer with HR+ and HER2+ disease by Genetically Inferred Ancestry.**

Associations between PRS$_{313}$ and HR+ and HER2+ in American women of African (AA), European (EA), East Asian American (EAA), and Hispanic (HL), based on genetically inferred ancestries. The top table shows that the PRS$_{313}$ is only associated with HR+ disease in European American patients (p-value = 0.0002) but not other ancestries. In contrast, the bottom table shows that the PRS$_{313}$ is associated with HER2+ disease but only for Hispanic American patients (p-value = 0.002). Even after accounting for age at diagnosis, the PRS$_{313}$ remains significantly predictive of HER2+ disease in Hispanic Americans (OR, 2.52; 95 CI, 1.35-4.70; p-value = 0.003).

|     | HR- | HR+ | OR | Lower CI | Upper CI |
| --- | --- | --- | --- | --- | --- |
| AA  | 15  | 51  | 0.71 | 0.36 | 1.40 |
| EA  | 176 | 768 | 1.42 | 1.16 | 1.64 |
| EAA | 22  | 114 | 1.38 | 0.41 | 1.20 |
| HL  | 33  | 106 | 0.70 | 0.91 | 2.20 |

|     | HER2- | HER2+ | OR | Lower CI | Upper CI |
| --- | --- | --- | --- | --- | --- |
| AA  | 55  | 11  | 2.26 | 0.96 | 5.33 |
| EAS | 814 | 130 | 0.95 | 0.79 | 1.15 |
| EAA | 110 | 26  | 1.02 | 0.62 | 1.66 |
| HL  | 121 | 18  | 2.47 | 1.39 | 4.41 |